# Fully Memristive Spiking-Neuron Learning Framework and Its Applications on Pattern Recognition and Edge Detection

Zhiri Tang[a,b], Yanhua Chen[c], Shizhuo Ye[a], Ruihan Hu[a], Hao Wang[a], Jin He[a], Qijun Huang[a], Sheng Chang[a*]

[a] School of Physics and Technology, Wuhan University, Wuhan, 430072, China
[b] Department of Computer Science, City University of Hong Kong, Hong Kong, 999077, China
[c] Department of Geography, The University of Hong Kong, Hong Kong, 999077, China

*Abstract*—Fully memristive neuron learning framework, which uses drift and diffusion memristor models to build an artificial neuron structure, becomes a hot topic recently with the development of memristor. However, some other devices like resistor or capacitor are still necessary in recent works of fully memristive learning framework. Theoretically, if one neuron is built by memristors only, the technique process will be simpler and learning framework will be more like biological brain. In this paper, a fully memristive spiking-neuron learning framework is introduced, in which a neuron structure is just built of one drift and one diffusion memristive models and spikes are used as transmission signals. The learning framework and spiking coding mode are simple and direct without any complicated calculation on hardware. To verify its merits, a feedforward neural network for pattern recognition and a cellular neural network for edge detection are designed. Experimental results show that compared to other memristive neural networks, processing speed of the proposed framework is very high, and the hardware resource is saved in pattern recognition. Further, due to the dynamic filtering function of diffusion memristor model in our learning framework, its peak signal noise ratio (PSNR) is much higher than traditional algorithms in edge detection.

## 1. Introduction

The memristor, which was postulated by L.O. Chua [1] in 1971, is the fourth basic circuit element alongside with the resistor (R), inductor (L), and capacitor (C). After HP Labs made memristors by TiO2 [2] in 2008, various memristors are realized. In general, memristors can be divided into two main types [3]: drift memristor and diffusion memristor. Correspondingly, a regular neuron in the human brain always consists of two main parts [4]: axon and dendrite. The function of axon, which is used as a channel for transmitting spiking signals [5], is similar to drift memristor [6], and function of dendrite, which is used as a receptor that receives spiking signals from the previous neuron [7], is the same as diffusion memristor theoretically [8].

Based on the above analogy, the fully memristive neuron, which uses drift and diffusion memristor as 'axon' and 'dendrite' in our artificial neuron structure respectively, has become a candidate for simulating biological neurons and realizing some brain-like learning rules [9-11]. Furthermore, benefiting from merits of memristor, such as relatively simple technique process, and being closer to biological brain neuron [12], memristive neurons can form memristive neural networks, which have immense potential applications in many areas including high performance computing [13-15], pattern recognition [16-18], and edge detection [19-21]. Some state-of-the-art works about memristive neural networks for pattern recognition, such as unsupervised memristive crossbar structure [22] and feedforward memristive neural network [23], focused on transforming image into other signals. In [22] and [23], memristors are used as a part of the entire systems and work with other modules. Some other state-of-the-art works for edge detection were mainly memristive cellular neural networks [24, 25] and multilayer perceptron with memristor crossbar [26]. Similarly, memristors in [25] and [26] work as an important part rather than an individual module to processing signals, which limits the processing speed of the entire systems and bring some difficulties to the practical production.

Inspired by above, some recent works focused on how to build a fully memristive learning framework. The research on fully memristive neural networks built a memristive crossbar by drift memristors, diffusion memristors, and capacitors [22] while another research used memristors and metal-oxide-semiconductor field effect transistors (MOSFET) [27]. From the above one can see that some other devices are still needed. Originally, drift and diffusion memristors should be merged into a basis neuron structure naturally and this neuron should use spiking signals as the information [28] which is transmitted between neurons. Hence, the novel fully memristive learning framework needs to be developed aiming to various applications.

However, as novel devices, memristor has some limitations including incompatibility with mainstream hardware [29] and parameter fluctuations [30], which limits the applications of memristive systems. If memristor models in software platform are used to build the entire memristive systems, the merits of in-memory computing benefiting from memristor, including high processing speed and low resource occupancy, can't be verified. Hence, implementing memristive systems or learning framework via mature hardware platforms [31] is one of the best choices to explore the applications of memristive systems. Some standard integrated circuit technologies [32, 33], including field programmable gate arrays (FPGA) and application specific integrated circuits (ASIC), are convenient for implementing memristive systems benefiting from the merits of these hardware tools such as high processing speed, robustness, and low power consumption [34]. For instance, TrueNorth, which was first proposed by IBM and designed a neuromorphic architecture based on spike-timing prototyped, has quite low power consumption and high speed for communication [35].

* Corresponding author
[a] E-mail: changsheng@whu.edu.cn (Sheng Chang)

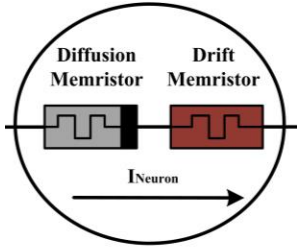

**Fig. 1.** Basic structure of artificial neuron including drift and diffusion memristor models.

Based on the above, this paper designs a fully memristive spiking-neuron learning framework using drift and diffusion memristor models only and applies it in pattern recognition and edge detection on hardware. The main contributions can be summarized as follows:

a. A fully memristive spiking-neuron learning framework is designed. Only drift and diffusion memristor models are used to build a spiking-neuron structure without any other electron devices. This framework only needs the most basic memristive characteristics and uses spiking information in the learning process directly, which is friendly and universal for various memristor models.

b. A fully memristive feedforward neural network is designed based on the spiking-neuron, which can achieve high pattern recognition accuracy. Due to the simple and direct spiking coding mode and learning framework without any complicated calculation on hardware, it can reduce the hardware resource occupancies significantly and improve the processing speed quite efficiently.

c. A fully memristive cellular neural network is designed for edge detection based on the spiking-neuron. As the diffusion memristor model in the learning framework has dynamic filtering function, it can form a system with high peak signal to noise ratio (PSNR) and has a good performance on edge detection including natural images and remote sensing images.

## 2. Fully Memristive Spiking-Neuron Learning Framework

In this section, a fully memristive spiking-neuron learning framework is introduced, including how the drift and diffusion memristor models connect, how spiking information transmit between neurons and why this framework can mimic biological neurons better.

*2.1. Basic Structure of Spiking-Neuron*

Current researches show that drift and diffusion memristor models can mimic characteristics of artificial neuron well [22]. Hence, the memristive spiking-neuron structure obeys this law, which is shown as Fig. 1.

When the hold time of input spikes are the same, we can use the basis diffusion memristor model [36] as follow:

$$\begin{cases} I_{output}(t) = U_{input}, & U_{input} > Threshold \\ I_{output}(t) = 0, & U_{input} \leq Threshold \end{cases} \quad (1)$$

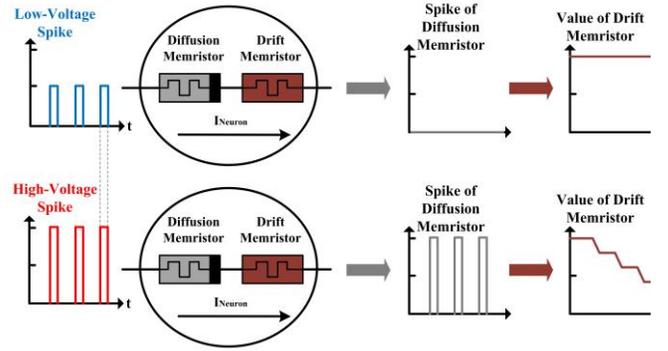

**Fig. 2.** Unsupervised learning algorithm of fully memristive spiking-neuron model.

The diffusion memristor model can be regarded as a threshold switch to determine whether a spike from the previous neuron was worth firing back.

For the drift memristor in the traditional HP model, the relationship of the voltage at the two ends of the memristor and its current is

$$U(t) = [R_{off} - (R_{off} - R_{on})\mu_v \frac{R_{on}}{D^2} \int_{-\infty}^{t} I(t)dt]I(t) \quad (2)$$

So the drift memristor value is

$$M_{drift}(t) = R_{off} - (R_{off} - R_{on})\mu_v \frac{R_{on}}{D^2} \int_{-\infty}^{t} I(t)dt = k_1 - k_2 \int_{-\infty}^{t} I(t)dt \quad (3)$$

where $k_1 = R_{off}$ and $k_2 = (R_{off} - R_{on})\mu_v R_{on}/D^2$.

From above, the value of drift memristor linearly dependents on the amount of electricity (Q) flowing through it. The value of drift memristor is the reciprocal of synaptic weight [6] in neuron, which can be increased or decreased during training process.

*2.2. Unsupervised Learning Algorithm*

In our learning framework, the diffusion memristor model works as a threshold switch in neuron while the drift memristor model serves as synaptic weight, which is shown as Fig. 2.

During the training process, high-voltage or low-voltage spikes are inputted to this neuron structure. When the input is low, the diffusion memristor won't response. When the input is high, the diffusion memristor will fire backward and the output spike of the diffusion memristor is the same as the input spike, which will decrease the value of the drift memristor. Since the value of drift memristor is the reciprocal of synaptic weight in a neuron, the synaptic weight of this neuron will increase receiving the high-voltage spike.

$M_{drift\ i+1}$ is the value of the drift memristor after one input spike, which is described as:

$$M_{drift\ i+1} = \begin{cases} k_1 - k_2 \dfrac{U_{input}}{M_{drift\ i}} t_{spike}, & U_{input} > Threshold \\ M_{drift\ i}, & U_{input} < Threshold \end{cases} \quad (4)$$

where $M_{drift\ i}$ is the value of the drift memristor before the input spike. $t_{spike}$ is the duration of this input voltage spike, which is very short so that the change value of the drift memristor during the spike time can be ignored.

Considering the spikes are continuously inputted during training process, the synaptic weight of the neuron with high input voltage will increase along with increasing number of input high-voltage spikes. The synaptic weight of the neuron with low input voltage won't have any change. After training, the synaptic weights with high-voltage spikes are much higher than those with low-voltage spikes.

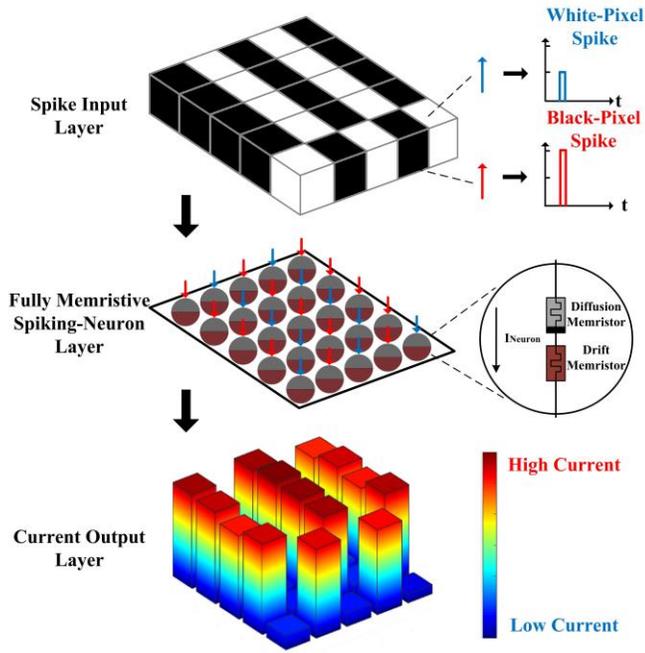

**Fig. 3.** The architecture of fully memristive feedforward neural network.

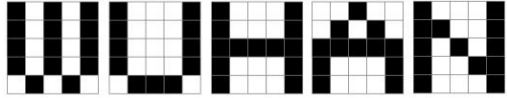

**Fig. 4.** 5*5 alphabet recognition dataset 'WUHAN'

In test process, the output currents from the drift memristors vary after training. The currents of drift memristors with the lower values, which means the higher synaptic weights, will be higher so that electrical currents prefer the shorter or thicker paths. Based on above, a self-competitive process is built naturally. The synaptic weight is only determined by the input voltage and it directly determines how much current it can 'grab'. Following the self-competitive process, the learning algorithm of this neuron structure is unsupervised.

In this framework, only basic drift and diffusion memristor models are employed. All the neuron structure needs are the most elementary characteristics of the memristor rather than special materials or structures. In other words, our learning framework is universal to various memristors.

## 3. Fully Memristive Feedforward Neural Network for Pattern Recognition

To verify the merits of our learning framework such as the simple connection mode and direct information representation by spiking signals and excluding complicated calculation on hardware, a fully memristive feedforward neural network for pattern recognition is designed with unsupervised learning process in this section. The experimental results show that the network has extremely high maximum clock frequency and low resource maintaining the high recognition accuracy.

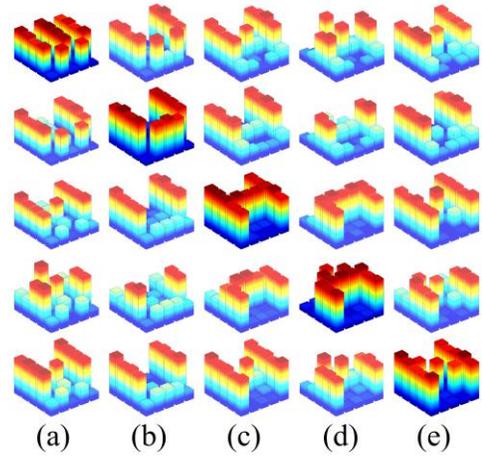

**Fig. 5.** Test results of 5*5 alphabet dataset 'WUHAN' with random noise (a) test results of alphabet 'W' (b) test results of alphabet 'U' (c) test results of alphabet 'H' (d) test results of alphabet 'A' (e) test results of alphabet 'N'.

### 3.1. Architecture of Network

The architecture of this network is shown as Fig. 3. The overall feedforward neural network is divided into three layers: the spike input layer, the fully memristive spiking-neuron layer and the current output layer. First, the input images are coded into spikes, in which the white and black pixels are coded into the low and high voltage spikes, respectively.

Second, in the fully memristive spiking-neuron layer, each pixel needs one fully memristive spiking-neuron. Given an example of an image with 5*5 pixels, the number of neurons is 25. As mentioned above, the diffusion memristor is used as a threshold switch which is higher than the white-pixel spike and lower than the black-pixel spike. Hence the fully memristive spiking-neurons corresponding to black-pixels will 'open', which means the values of the drift memristor models in these neurons will decrease and the synaptic weights of these neurons will increase.

Third, in the current output layer, the output current values depend on the values of the drift memristors. If a pixel is black, the drift memristor corresponding to this pixel will decrease during the training process. Then its output current value will be higher than the memristors corresponding to white-pixels.

In this network, the number of categories is the number of output spikes and the result of image classification can be obtained by comparing the sum of the output current corresponding to each category directly. The one with the highest total output current, which means the one with the lowest total value of the drift memristor, represents the correct category.

**Table 1**
Test resources of dataset 'WUHAN'

|  |  | Expectation | | | | |
| --- | --- | --- | --- | --- | --- | --- |
|  |  | W | U | H | A | N |
| **Experiment results** | W | 200 | 0 | 0 | 0 | 0 |
|  | U | 3 | 196 | 0 | 0 | 1 |
|  | H | 1 | 0 | 199 | 0 | 0 |
|  | A | 1 | 0 | 1 | 198 | 0 |
|  | N | 1 | 0 | 1 | 0 | 198 |

**Table 2**
Comparisons of maximum clock frequency and resource consumption

| Design | FPGA Platform | Maximum Clock Frequency (MHz) | Resource |
|---|---|---|---|
| DMS-ANN [37] | Cyclone II: EP2C70F672C6 | N/A | 3266 (LEs) |
| Our Network | Cyclone II: EP2C70F672C6 | 165.32 | 952 (LEs) |
| TC-Network [38] | Cyclone IV: EP4CE115 | 40.68 | 3031 (LEs) |
| Our Network | Cyclone IV: EP4CE115F23C9 | 96.67 | 951 (LEs) |
| HF-MNN [23] | Stratix V: 5SGXEA7N2F45C2 | 256.54 | 869 (in ALMs) |
| Our Network | Stratix V: 5SGXEA7N2F45C2 | 448.23 | 477 (in ALMs) |

*3.2. Experimental Results and Analysis*
A home-made 5*5 alphabet recognition dataset 'WUHAN', which is shown as Fig. 4, is used to test the performance of our fully memristive feedforward neural network on Intel Field Programmable Gate Array (FPGA). Intel FPGA brands include MAX, Cyclone, Arria, and Stratix FPGAs, in which the Cyclone and Stratix are common FPGAs for research and experiment analysis. Cyclone are mainly low-level FPGAs, which have relatively lower processing speed. Most Stratix are high-level FPGAs aiming to better hardware optimization and higher maximum frequency. Through Quartus Prime software, the compiling results are presented in flow summary including resource occupancy (LEs in Cyclone except Cyclone V, and ALMs in Cyclone V and Stratix) and maximum frequency (having difference between different FPGA brands to some extent). Hence, to verify the experimental results, the proposed algorithms need to be implemented on the same or similar FPGA platforms.

During one training period, five images including alphabet 'W', 'U', 'H', 'A', and 'N' are input to the network. The entire training process only needs three training periods. The test results with 20% random noise on standard images are shown as Fig. 5. The five images of Fig. 5(a) are the output currents corresponding to the five types as the input test image is randomly noised 'W'. It is clear that the first output image has the highest current as it portrays the alphabet 'W', which is the correct classification result. Likewise, the noised alphabet 'U', 'H', 'A', and 'N' are correctly classified in Fig. 5(b), (c), (d), and (e). 200 sets of images with 20% random noise, totally 1000 images, are tested. The test results are shown as Table 1. From the table, only 9 images are identified into the wrong classifications among the 1000 images and the accuracy is over 99%, which shows good classification performance.

To verify the network's performance, the FPGA design of our network is implemented by Quartus Prime software and compared with an artificial neural network using a digital memristor simulator (DMS-ANN) [37], a two-compartment memristive network (TC-Network) [38], and a hardware friendly memristive neural network (HF-MNN) [23] with the same network scale on the same platform, respectively. Because the TC-Network didn't indicate specific FPGA platform, we select the slowest FPGA in the same series as the TC-Network, which is Cyclone IV: EP4CE115F23C9 (The relative processing speed is up to the last number of FPGA, which is "9" for EP4CE115F23C9. The larger the number is, the lower the processing speed is in the same FPGA series). From the results in Table 2, the maximum clock frequency of our network in Stratix V is 448.23MHz so that the training of our network needs only $3/(448.23 \times 10^6) \approx 6.69ns$. Our network gains extremely high maximum clock frequency with low resource consumption, which is derived from its brief neuron connection mode and efficient spiking coding and the learning framework excluding any complicated calculation on hardware. Specially, our memristive learning framework eliminates other devices normally needed by memristive neural networks, so the hardware resource is saved. Since the spiking information is directly processed by fully memristive neuron structure ignoring additions and multiplications, processing speed is improved significantly.

**4. Fully Memristive Cellular Neural Network for Edge Detection**

Since the diffusion memristors in our learning framework are regarded as dynamic filtering thresholds, a fully memristive cellular neural network for edge detection is designed to show our framework's merits on anti-noise in this section. Besides it can achieve high PSNR with noised images, its edge detection results on the complex objects are still quite good.

*4.1. Architecture of Network*
The basic idea comes from a phenomenon, which is known as that the human eye's strong adaptability to brightness varies and this adaptability depends on the current brightness. That means human eye can't notice the brightness changes until the change goes beyond the specific threshold [39]. Utilizing this theory, edge detection methods can be developed. Some researchers have concluded the human eye's detection thresholds for red, green, and blue, respectively, based on current brightness values, which is shown as follow:

$$\Delta r = \begin{cases} 12734e^{-0.1494r}, & 0 \le r < 38 \\ 5397e^{-0.1015r}, & 38 \le r < 60 \\ 127300e^{\left[(0.07r-7.55)e^{0.026r}-0.089r\right]}, & 60 \le r < 97 \\ 8.4569, & 97 \le r \le 255 \end{cases} \quad (5)$$

$$\Delta g = \begin{cases} 12734e^{-0.1494g}, & 0 \le g < 38 \\ 5107.5e^{-0.1015g}, & 38 \le g < 60 \\ 120470e^{\left[(0.07g-7.55)e^{0.026g}-0.089g\right]}, & 60 \le g < 97 \\ 8.003, & 97 \le g \le 255 \end{cases} \quad (6)$$

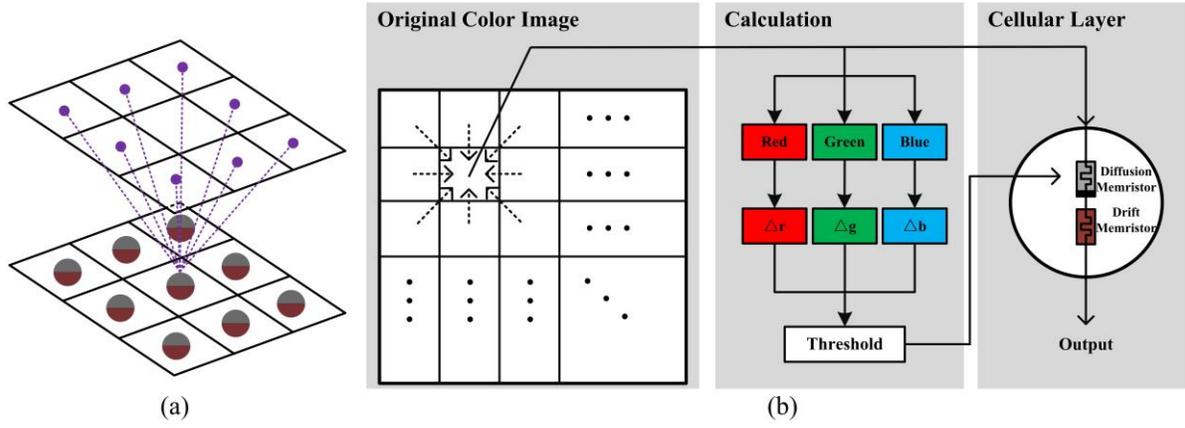

**Fig. 6.** The architecture of fully memristive cellular neural network (a) network connection mode (b) network flow.

$$\Delta b = \begin{cases} 12734e^{-0.1494b}, & 0 \leq b < 38 \\ 9101.6e^{-0.1015b}, & 38 \leq b < 60 \\ 241750e^{\left[(0.07b-7.55)e^{0.026b}-0.089b\right]}, & 60 \leq b < 97 \\ 14.2663, & 97 \leq b \leq 255 \end{cases} \quad (7)$$

where $r$, $g$, and $b$ are pixel values of red, green, and blue colors respectively, whose value ranges are from 0 to 255. $\Delta r$, $\Delta g$, and $\Delta b$ are respectively the minimum changes of red, green, and blue that human eyes can recognize to color images. However, since the proportion of the primary colors varies, some research [40] proposed a threshold $T_{i,j}$ for color images as follow:

$$T_{i,j} = \omega_r \Delta r_{i,j} + \omega_g \Delta g_{i,j} + \omega_b \Delta b_{i,j} \quad (8)$$

which has $\omega_r = 0.114$, $\omega_g = 0.587$, and $\omega_b = 0.299$.

According to the above, our network architecture is shown as Fig. 6. In the process of edge detection, the minimum changes of the three colors are calculated first according to the target pixel, and then the total thresholds are set. These thresholds are the switch thresholds of the diffusion memristors in cellular layer.

The eight pixels around the target pixel are compared with the target pixel in turn. If the difference value exceeds the threshold value of the diffusion memristor, it will fire backward and the value of the drift memristor will decrease. If not, the diffusion memristor won't fire and the value of the drift memristor won't change. The edge detection process is completed when all eight pixels are compared. After that, the same voltages are added to each neuron module in the cellular layer directly, just as we do above in pattern recognition. Finally, the obtained output current can form a diagram, which is the output image after edge detection.

*4.2. Experimental Results and Analysis*

First, the classic edge detection image 'Lena' with 512*512 pixels is used to test the anti-noise performance of our fully memristive cellular neural network as shown in Fig. 7(a). The diffusion memristors in the fully memristive cellular neural network act as dynamic filtering thresholds, which can increase the network's PSNR theoretically. The Gauss noise with mean 0 and variance 0.1 is added to the 'Lena' and the test results are given as Fig. 7(b). Compared to traditional algorithms such the Sobel, Robert, Prewitt, LoG, and Canny, the PSNR of our network is over 21dB

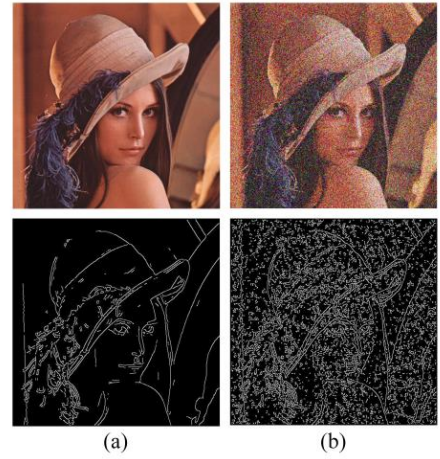

**Fig. 7.** Edge detection results of 'Lena' (a) without noise (b) with noise

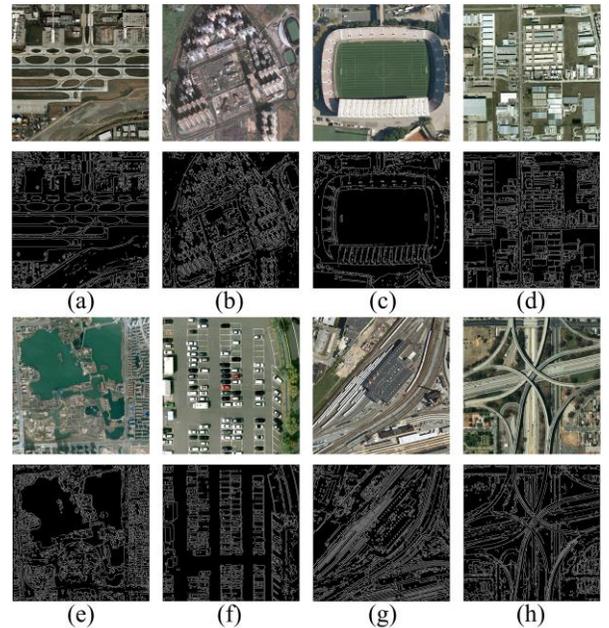

**Fig. 8.** Edge detection results of remote sensing images (a) airport (b) commercial district (c) football field (d) industrial district (e) park (f) parking lot (g) railway station (h) viaduct

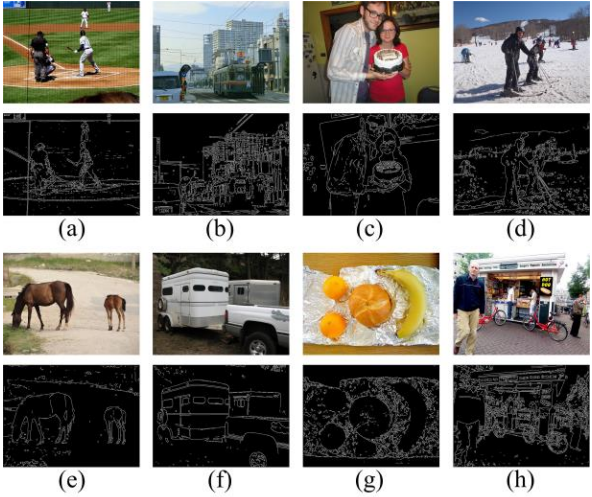

**Fig. 9.** Edge detection results of COCO dataset

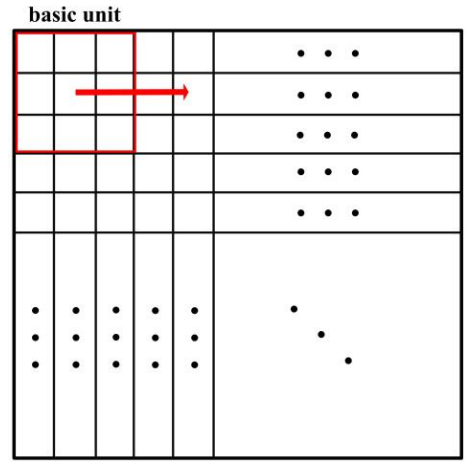

**Fig. 10.** Overview of a basic hardware unit (slide window) to load and calculate the images

**Table 3**
Comparisons of PSNR among some edge detection algorithms

| Test image | Sobel (dB) | Robert (dB) | Prewitt (dB) | LoG (dB) | Canny (dB) | Our Network (dB) |
|---|---|---|---|---|---|---|
| Lena | 13.6548 | 13.6593 | 13.6822 | 10.0023 | 8.5448 | 21.1933 |

**Table 4**
Anti-noise performance of urban remote sensing images

|  | Airport | Commercial District | Football Field | Industrial District | Park | Parking Lot | Railway Station | Viaduct |
|---|---|---|---|---|---|---|---|---|
| PSNR(dB) | 20.0930 | 19.6833 | 19.3321 | 19.5697 | 19.5116 | 19.4885 | 20.0409 | 20.1920 |

**Table 5**
Anti-noise performance of COCO dataset

|  | person | vehicle | outdoor | animal | accessory | sports |
|---|---|---|---|---|---|---|
| PSNR(dB) | 22.0184 | 21.9896 | 21.1781 | 21.5801 | 21.9903 | 22.1395 |
|  | kitchen | food | furniture | electronic | appliance | indoor |
| PSNR(dB) | 22.3166 | 22.0406 | 21.6837 | 23.8230 | 21.7464 | 20.9117 |

**Table 6**
Hardware performance of edge detection

| Test Images | FPGA Platform | Resource | Throughput |
|---|---|---|---|
| Lena | Stratix V: 5SGXEA7N2F45C2 | 221 (in ALMs) | 1,723 images/s |
| WHU-RS19 | Stratix V: 5SGXEA7N2F45C2 | 221 (in ALMs) | 1,253 images/s |
| COCO | Stratix V: 5SGXEA7N2F45C2 | 221 (in ALMs) | 1,469 images/s |

and the highest of other traditional algorithms is below 14dB as shown in Table 3.

To illustrate performance of the proposed network in complex images, a remote sensing image dataset with 600*600 pixels from Google satellite images, WHU-RS19 [41], and a complex natural image dataset, COCO [42], are tested.

WHU-RS19 is very suitable to verify a method's anti-noise capacity, because the edge information of the urban remote sensing images are complex and significant, which are always with various noises. Specifically, we test eight different types of the urban scenes, including the airport, commercial district, football field, industrial district, park, parking lot, railway station and viaduct as shown in Fig. 8.

Due to the dynamic filtering function of the diffusion memristor models in our network, the edge detection results of the urban remote sensing images are very good. In Fig. 8(a), the edges of all the runways and terminals, even the location of the airplanes are very clear. In Fig. 8(b), the edges of both the streets and buildings are distinct. In Fig. 8(c), the edges of the stadium frames and white lines on the court are distinguished and the green stripes on the court are filtered out. In Fig. 8(d), the edges of all the farmlands and industrial buildings are obvious. In Fig. 8(e), the edges of the lakes in the park are clear which is helpful to distinguish the boundary between the lake and the road clearly. In Fig. 8(f), the edges of all parking lines and vehicles are very distinct which is helpful for people to distinguish whether there is a vehicle in this parking space or not. In Fig. 8(g), the edges of all the trains are appeared and even the number of the trains in the railway station can be counted. In Fig. 8(h), the edges of all the viaducts are very clear and it can be distinguished which viaduct is higher and which is lower. To test the anti-noise performance of urban remote sensing images, the same Gauss noise on the 'Lena' is added to these eight image types. The PSNR of noised images ranges from 19 to 22, which shows a good anti-noise performance as shown in Table 4.

As one of the most common datasets in computer vision, COCO is also tested to verify the performance of the proposed memristive cellular neural networks. COCO can be generally divided into 12 types including person, vehicle, outdoor, animal, accessory, sports, kitchen, food, furniture, electronic, appliance, and indoor. Similarly, the edge detection performance of some images in COCO is shown in Fig. 9 and anti-noise performance is shown in Table 5, from which the proposed framework has a good performance in various datasets.

Further, according to the proposed framework in Fig. 6, 9 pixels are used as a basic unit. Hence, the basic hardware units with 9 fully memristive spiking neurons can be regard as a slide window to load and calculate the images, which is shown in Fig. 10. The hardware performance in the above images is shown in Table 6, from which it is verified that the proposed framework has quite high throughput with only few hardware resource occupancies.

## 5. Conclusions

In this paper, a fully memristive spiking-neuron learning framework, which uses diffusion and drift memristor models only and applies spiking as transmission signals, is presented. Due to its simple structure and fluent spiking information flow without any complicated calculation on hardware, the learning framework has many merits including high maximum clock frequency, low resource and high anti-noise performance. These merits are verified by some applications including the fully memristive feedforward neural network for pattern recognition and the fully memristive cellular neural network for edge detection.

Further, the input signals of the proposed neuron structure have the same form as output signals, so it can be regarded as a recyclable structure. In the future, the proposed memristive spiking neuron has the potential to build a multilayer neural network to handle some other complex tasks including processing time series [43] or EEG signals [44]. We hope this idea can give an inspiration for the developments of the fully memristive neuron and neural networks, and other neuromorphic algorithms.

## Acknowledgements

This work was supported by the National Natural Science Foundation of China (61874079 and 61574102), the Fundamental Research Fund for the Central Universities, Wuhan University (2042017gf0052), the Wuhan Research Program of Application Foundation and Frontier Technology (20180104010111289), and the Luojia Young Scholars Program. Part of calculation in this paper has been done on the supercomputing system in the Supercomputing Center of Wuhan University.